\documentstyle[subeqn,12pt]{article}
\def\be{\begin{equation}}
\def\ee{\end{equation}}
\def\beq{\begin{eqnarray}}
\def\eeq{\end{eqnarray}}

\begin{document}
%\onehalfspace
\begin{flushright}
TIFR/TH/96-28
\end{flushright}
\bigskip
\bigskip
\begin{center}
{\large{\bf Theory of the Marginal Fermi Liquid spectrum in the 
local treatment of the large-U Falicov-Kimball Model.
 }}\\
\bigskip
\bigskip
\bigskip
\medskip
{\bf Mukul S. Laad }\\
\bigskip
{\it Tata Institute of Fundamental Research, \\
Homi Bhabha Rd, Colaba, Bombay 400 005, India} \\
\bigskip
\bigskip
%{\Large{\bf Abstract}}\\
\end{center}
\bigskip

\begin{abstract}
 The Marginal Fermi Liquid (MFL) hypothesis proposed by 
Varma $ {\it et.al} $ is derived from a consistent treatment 
of local spin fluctuations in the Falicov-Kimball (FK) Model.  
Within an infinite-dimensional mean-field approach, which is 
exact for this model in $ d = \infty $, a marginal Fermi liquid 
susceptibility and single particle self-energy are 
obtained near half-filling, and a Fermi liquid phase is recovered 
sufficiently away from half-filling, in agreement with indications 
from experiments which probe the normal state of cuprates as a 
function of doping.      
\end{abstract}

\newpage

The discovery of high-T$_c$ superconductors in $Cu-O$ based compounds
has led to an upsurge of theoretical work concerning the unusual
normal state properties of these materials, which appear not to
conform to the framework of the Landau Fermi liquid theory [1, 2]. A
way of unifying the diverse anomalies observed in experiment was
proposed by Varma and coworkers [2], who suggested a phenomenological
ansatz for the spectrum of charge and spin fluctuations. For low
frequencies $\omega \ll v_Fq$ this marginal-Fermi-liquid (MFL) ansatz
is
\be
Im \chi_{\rho, \sigma} (\omega) \sim \left \{ \matrix{
- \rho (0) \frac{\omega}{T} , & \omega \ll T , \cr
- \rho (0) , & T \ll \omega \ll \omega_c} \right.
\ee
where $\omega_c$ is a cut-off energy. The s.p. self-energy $\sigma
(\omega) \sim \omega \ell n \omega \pm i | \omega |$, as a consequence
of (1), and this reconciles the unusual normal state anomalies with
the existence of the Luttinger Fermi surface.

As emphasized in [2], the singularities in (1) are in the frequency
dependence; the momentum dependence is assumed smooth. In principle,
an exact solution of certain strongly correlated models in $d =
\infty$, where local fluctuations are treated exactly, should lead to
the above spectrum. Furthermore, if these singularities do not depend
on any special symmetries which are lost in the lattice problem, they
are likely to survive in the lattice problem.

Varma et. al [2, 3] have solved the multiband Hubbard model within the
impurity approximation to obtain the MFL form for the local
susceptibilities. However, the MFL theory has not been able to account
for the $T$ and doping ($x$) dependence of the Hall constant $R_H$.
Recently, Mahesh et. al [4] have computed the $T$ and $x$ dependence
of $R_H$ by numerical diagonalization of the one-band Hubbard model on
finite-sized clusters. They were able to account for the anomalous
behavior of $R_H$. However, a proper description of {\it all} the
anomalous features has not been possible, and the extension of the
Luttinger liquid [5] concept of Anderson to two dimensions is not
clear, unless the small momentum forward scattering couplings become
singular [6].

In this letter, we show that a consistent treatment of local
fluctuations in the doped, large $U$ Falicov-Kimball model leads to
the Eqn. (1). We utilize the exact solution of the infinite dimensional
effective Falicov-Kimball (FKM) model [7]. An explicit analytical
calculation of the dynamical spin susceptibility $\chi_\sigma
(\omega)$ is consistent with the MFL form.

In this paper, the Falicov-Kimball model in 2d in the large $U$ limit,
\be
H = - t \sum_{<ij>} (c^\dagger_{i} c_{j} + h.c) + U
\sum_i n_{ic} n_{id} - \mu \sum_{i}(n_{ic} + n_{id})
\ee
is proposed as an effective model capable of describing the
anomalous properties of oxide superconductors in their normal state.
In this model, $t$ and $U$ should be understood as effective
parameters which are determined by comparison of the low-energy
spectra of (2) with that of a more realistic three-band model [2,8]. 
The proposed FKM bears some similarity to the effective model 
(eqs (12)-(14) of ref. [2]).  However, since the authors of ref. [2] 
solve an impurity model, they require fine-tuning of parameters to 
reach the critical point.  
Since we perform a lattice calculation exact in $ d = \infty $, the 
critical behavior survives for a finite range of filling, as the authors 
of ref. [2] anticipate.
In this FKM [7],
the $ d $ holes are immobile. This means that
$[n_{id}, H] = 0 \forall i$; hence, the model has an exact
local $U(1)$ symmetry.  We notice that the model eqn (2) is different 
from the usual Hubbard model, which has a global U(1) symmetry 
associated with ${\it total}$ fermion number conservation. 
As we shall see, it is this local symmetry
which leads to the breakdown of Fermi liquid theory in our approach. 

Since we are interested in the nontrivial local dynamics, we consider
an auxiliary impurity model in which the $ d $  hole does not
hybridize with the ``conduction electrons''. This impurity model
hamiltonian is the $d = \infty$ counterpart of Eqn. (3), and reads
\be
H = \sum_k \varepsilon_k c^\dagger_{k} c_{k} + t \sum_k
(e^{ik \cdot R_i} c^\dagger_{i} c_{k} + h.c) + U
n_{ic} n_{id} - \sum_{i} \mu (n_{ic} + n_{id})
\ee
where $i$ represents the impurity site, and $k$ labels the
``conduction electrons''. We are interested in the non-trivial local
dynamics; hence, we compute the local s.p and two-particle propagators
exactly. To compute the s.p Green function, we start with an equation
of motion for it.
\beq
(i\omega_\ell + \mu)G^{c}_{ii} (i\omega_\ell) &=& \frac{1}{2\pi} +
\sum_k t_k G^{c}_{ki} (i\omega_\ell) + U \langle\langle
n_{id} c_{i};
c^\dagger_{i}\rangle\rangle_{i\omega_\ell} \nonumber \\ [3mm]
(i\omega_\ell + \mu - U) \langle\langle n_{id} c_{i};
c^\dagger_{i}\rangle\rangle_{i\omega_\ell} &=&
\frac{<n_{id}>}{2\pi} + \sum_k t_k \langle\langle n_{id}
c_{k}; c^\dagger_{i}\rangle\rangle_{i\omega_\ell}\nonumber\\ [3mm]
(i\omega_\ell - \varepsilon_k) G^{c}_{ki} (i\omega_\ell) &=& t_k
G^{c}_{ii} (i\omega_\ell) \nonumber \\ [3mm]
(i\omega_\ell - \varepsilon_k) \langle\langle n_{id}
c_{k}; c^\dagger_{i}\rangle\rangle_{i\omega_\ell} &=& t_k
\langle\langle n_{id} c_{i};
c^\dagger_{i}\rangle\rangle_{i\omega_\ell}
\eeq
solving for $G^{c}_{ii} (i\omega_\ell)$ yields (for a Lorentzian
unperturbed DOS with a half\-width $\Delta$,  
$\rho_o (z) = (\Delta / \pi) (z^2 + \Delta^2)^{-2}$)
\subequations
\be
G^{c}_{ii} (i\omega_\ell) = \frac{1}{2\pi}
\left[\frac{1-n_{d}}{i\omega_\ell + \mu + i \Delta  
sgn \omega_\ell} + \frac{n_{id}}{i\omega_\ell + \mu - U + i\Delta
sgn \omega_\ell}\right]
\ee
with the self-energy
\be
\Sigma_{c} (i\omega_\ell) = - \mu + Un_{d} + \frac{U^2
n_{d} (1 - n_{d})}{i\omega_\ell + \mu - U (1 - n_{d})
+ i \Delta sgn \omega_\ell}
\ee
\endsubequations
Also, it is easily seen that
\be
\langle\langle n_{id} c_{i}; c^\dagger_{i}
\rangle\rangle_{i\omega_\ell} = \frac{n_{d}}{2\pi}
\frac{1}{i\omega_\ell + \mu - U + i\Delta sgn \omega_\ell}
\ee
The s.p and the two-particle local spectral densities are
\be
\rho_{c} (i\omega_\ell) = \frac{\Delta}{2\pi^2}
\left[\frac{1-n_{d}}{(i\omega_\ell + \mu)^2 + \Delta^2}
+ \frac{n_{d}}{(i\omega_\ell + \mu - U)^2 + \Delta^2}\right]
\ee
and
\be
\rho^{(2)} (i\omega_\ell) = \frac{\Delta}{2\pi^2} \frac{n_{d}}
{(i\omega_\ell + \mu - U)^2 + \Delta^2}
\ee
>From Eqns. (7) and (8), it is clear that the low-energy spectrum is a
superposition of s.p and two-particle states. It is precisely the
resonant scattering between these states that leads to the breakdown
of FLT.

To proceed with the derivation of the MFL spectrum, we compute the NMR
relaxation rate, which is related to the low-frequency dynamical local
spin susceptibility via
\beq
\frac{1}{T_1} &=& - T \lim_{\omega to 0} \sum_{\vec q} \frac{\chi''
(\vec q, \omega)}{\omega} \ , \ {\rm for} \ \omega \ll T \nonumber \\ [3mm]
&=& - \sum_{\vec q} \chi'' (\vec q, \omega) , \ \ \ \ \ \ \ \ \ {\rm
for} \ T \ll \omega .
\eeq
where $\chi'' (\omega)$ is the imaginary part of the dynamical spin
susceptibility. We also have [9]
\beq
\frac{1}{T_1} &=& \frac{A^2}{2\hbar N} \sum_i \int^{+\infty}_{-\infty}
\langle T_\tau [S^+_i (\tau) S^-_i (0)] \rangle e^{i\omega\tau} d\tau  
\bigg|_{\omega \rightarrow 0} \nonumber \\ [3mm]
&=& \frac{A^2}{2\pi\hbar\Delta N} \sum_i \langle S^+_i S^-_i \rangle
\eeq
so that the task reduces to calculating (exactly within $d = \infty$)
the local, transverse spin correlation function. This can be computed
easily once $\rho^{(2)} (i\omega_\ell)$ is known, by using the
identity
\be
\langle S^+_i S^-_i \rangle = \frac{n_d}{2} - \langle
n_{ic} n_{id} \rangle
\ee

But $D = \langle n_{ic} n_{id}\rangle$, the average
number of doubly occupied sites is computed directly from the
two-particle spectral density via a Matsubara sums
\be
\langle n_{ic} n_{id} \rangle = - \frac{1}{\beta}
\sum_\ell \rho^{(2)} (i\omega_\ell) e^{-\omega_\ell (-0)}
\ee

Direct evaluation by use of Eqn. (9) yields
\beq
\langle n_{ic} n_{id}\rangle &=& \frac{n_{d}}{\pi}
\left[ \frac{\pi}{2} + \tan^{-1} \left(\frac{\mu - U}{\Delta}\right)
\right] \ , \ \ k_BT \ll \Delta \nonumber \\ [3mm]
&=& \frac{n_{d}}{\pi} \left[\frac{\pi}{2} + \tanh \left\{\beta
\left(\frac{\mu - U}{2}\right)\right\}\right] \ , \ \ k_BT \gg \Delta
\eeq
Hence from Eqns. (12) and (14), we get
\beq
\langle S^+_i S^-_i \rangle &=& \frac{n_{d}}{\pi}
\tan^{-1} \left(\frac{Un_{d}}{\Delta}\right)
\ , \ \ k_BT \ll \Delta \nonumber \\ [3mm]
&=& \frac{n_{d}}{\pi}  \tanh \left(\beta
\frac{Un_{d}}{2}\right) \ , \ \ k_BT \gg \Delta
\eeq
where we used the exact relation $\mu = U (1 - n_{d})$.
Comparison of Eqn. (10) with Eqn. (11) (after substitution of (15))
leads to
\beq
\chi'' (\omega) &=& - \frac{A^2}{2\hbar} \rho (0) n_{d}
\tan^{-1} \left(\frac{Un_{d}}{\Delta}\right)
\left(\frac{\omega}{T} \right) \ , \ \ \omega \ll T \nonumber \\ [3mm]
&=& - \frac{A^2}{4\hbar} \rho (0) n_{d} \ , \ \ \ \beta U \gg 1
\ , \ \ \ T \ll \omega
\eeq

This is precisely of the MFL form. A second-order perturbative
calculation gives $\Sigma (\omega) \sim \omega \ell n \omega - i
|\omega|$. Thus, a proper treatment of local spin fluctuations leads
to the MFL spectrum for the spin susceptibility as well as the
self-energy. It is important to notice that the MFL ansatz is a
statement about both the s.p self-energy as well as the
susceptibilities, and so approaches which compute only the self-energy
of the correct form [10] are inadequate.

Earlier attempts, besides those of [2, 3] have attempted to derive the
MFL spectrum by invoking a negative $U$ HM [11], or the Holstein model
[12]. The relevance of these models is questionable, since there are
no antiferromagnetic insulating or spin fluctuation dominated strange
metallic phases in these models. Our approach, which starts from the
2d, Falicov-Kimball model with large U, explicitly takes AFM local spin
fluctuations into account. In fact, noticing that the FKM in $d =
\infty$ is the recoilless x-ray edge problem [13], the local
``excitonic'', or the transverse spin-spin correlation function is
divergent at low energy near $n=1$,
\beq
\chi^{\pm \prime\prime} (\omega) &=& \int^\infty_0 \ d\tau e^{i\omega \tau}
\langle T_\tau [S^+ (\tau) S^- (0)] \rangle \nonumber \\ [3mm]
&\sim& | \omega |^{-\beta}
\eeq
with $\beta = \frac{2\delta}{\pi} - \left(\frac{\delta}{\pi}\right)^2$
and $\delta = \tan^{-1} \left(\frac{U}{\Delta}\right)$, the s-wave
phase shift at $\mu$. This leads to a soft, local spin fluctuation
mode at low energy, and it is precisely the coupling of the s.p part
of the spectrum to these soft modes which leads to the MFL ansatz.

The MFL ansatz reconciles the normal state anomalies observed in
cuprates with the existence of the Luttinger Fermi surface. We have
shown that the MFL ansatz can be derived from a Falicov-Kimball model. 
  Large $ U$
is crucial to the derivation, as is the filling factor. The MFL state
is unstable to a Fermi liquid (FL) phase for $n < n_c = 1 -
\frac{1}{\pi} cot^{-1} \left(\frac{U}{2\Delta}\right)$ [13]. With $U =
2\Delta$, for e.g., $n_c = 0.75$, i.e. the doping concentration of holes
$x = 0.25$, close to the experimental $x_c$ value beyond which the
anomalous behavior is suppressed [14].  At lower densities, effects 
of disorder
will introduce qualitatively new features.  A non-perturbative 
treatment including effects of disorder will be reported elsewhere.

We have not dealt with the superconducting phase in this paper;
however, a few remarks can be made. Due to large $U$, the on site 
pairing amplitude is severely suppressed, $\langle c^\dagger_{i}
d^\dagger_{i} \rangle \equiv 0$ (actually, because the Hamiltonian 
has an exact local U(1) symmetry, this is rigorously true, 
by Elitzur's theorem). This implies, upon Fourier
transforming, that
\be
\sum_{\vec k} \langle c^\dagger (\vec k) d^\dagger
(- \vec k) \rangle = 0
\ee

As pointed out by various authors [15, 16], the pair wave-function
should have lines or points on the FS at which it is zero, and hence
the pairing is {\it not} of the BCS variety. The gap function has
nodes at the same points where the pair wave function does. It is not
possible, from the above, to specify whether the symmetry of the SC
state is of the extended s-wave or the d-wave type. This is beyond the
scope of the present work.

In conclusion, we have shown that an exact treatment of the dynamical
spin fluctuations in the large $U$, one band Falicov-Kimball model leads to
the MFL phase near half-fillings and a FL phase sufficiently away from
$n = 1$.

\newpage

\noindent{\large{References}} \\

\begin{enumerate}
\item[{[1]}]
P.W. Anderson and Y. Ren, in ``Superconductivity - The Los Alamos
Symposium'', eds. K. Bedell, et. al, Addison Wesley.
\item[{[2]}]
C.M. Varma, P. Littlewood, S. Schmitt-Rink, E. Abrahams and A.E.
Ruckenstein, {\it Phys. Rev. Lett.} {\bf 63}, 1996 (1989); also C.
Sire, C.M. Varma, A.E. Ruckenstein and T. Giamarchi, {\it Phys. Rev.
Lett.} {\bf 72}, 15 (1994).
\item[{[3]}]
T. Giamarchi, Ph. Nozieres, C.M. Varma and A.E. Ruckenstein, {\it
Phys. Rev. Lett.} {\bf 70}, 3967 (1993).
\item[{[4]}]
P. Mahesh and B.S. Shastry, preprint.
\item[{[5]}]
P.W. Anderson, {\it Phys. Rev. Lett.} {\bf 64}, 1839 (1990).
\item[{[6]}]
see, for e.g., R. Shankar, in ``The Renormalization Group and its
Applications to Critical Phenomena'', Lecture Notes supplied during
Spring College in Condensed Matter on ``Quantum Phases'', 3 May -- 10
June, 1994, ICTP, Trieste, Italy.
\item[{[7]}]
Mukul S. Laad, {\it Phys. Rev.} {\bf B 49}, 2327 (1994); Y.M. Li and
N. d'Ambrumenil, {\it Phys. Rev.} {\bf B 49}, 6058 (1994); D.
Vollhardt, et. al, {\it Z. Phys.} {\bf B 91}, 329 (1993).
\item[{[8]}]
E. Stechel in [1], also, F.C. Zhang and T.M. Rice, {\it Phys. Rev.}
{\bf B37}, 3759 (1988).
\item[{[9]}]
R.M.White in ``Magnetism'', Publ. Springer Verlag, Vol. 32.
\item[{[10]}]
C. Ventura et. al, preprint, Centro Atomico, Bariloche and references
therein (1992).
\item[{[11]}]
C.M. Varma, {\it Int. J. Mod. Phys.} {\bf B3}, 12, 2083 (1989); in
fact, it has been fruitfully applied to bismuthates.
\item[{[12]}]
See for e.g., C.M. Srivastava, preprint (1991); the
Holstein model reduces to the -ve U HM after integrating out the
phonons in the limit $\omega_{ph} \rightarrow \infty$.
\item[{[13]}]
Q. Si, A. Georges and G. Kotliar, {\it Phys. Rev.} {\bf B46}, 1261 (1992).
\item[{[14]}]
B. Batlogg in [1].
\item[{[15]}]
A. J. Millis, H. Monien and D. Pines, {\it Phys. Rev.} {\bf B 42}, 167 
(1990).
\item[{[16]}]
S. Chakravarty, A. Sudbo, P.W. Anderson and S.P. Strong, {\it Science}
{\bf 261}, 337 (1993).
\end{enumerate}
\end{document}